\documentclass[debug]{epl}
\usepackage{epsfig,graphicx,times}

\begin{document}

\title{Mesoscopic Kondo Problem}

\author{R. K. Kaul \and D. Ullmo \and S.
  Chandrasekharan \and H. U.\ Baranger}
\shortauthor{R. K.\ Kaul \etal}
\institute{
   Department of Physics, Duke University, 
           Durham, NC 27708-0305.
}
\date{\today}


\maketitle 

\begin{abstract}
We study the effect of mesoscopic fluctuations on a magnetic impurity
coupled to a spatially confined electron gas with a temperature in the
mesoscopic range (i.e.\ between the mean level spacing $\Delta$ and
the Thouless energy $E_{\rm Th}$).  Comparing ``poor-man's scaling'' with
exact Quantum Monte Carlo, we find that for temperatures larger than
the Kondo temperature, many aspects of the fluctuations can be captured
by the perturbative technique. Using this technique in conjunction with
semi-classical approximations, we are able to calculate the mesoscopic
fluctuations for a wide variety of systems.  For temperatures smaller
than the Kondo temperature, we find large fluctuations and deviations
from the universal behavior.
\end{abstract}

The nature of many-body interaction effects at the nanoscale has been
of great interest recently.  Examples include, for instance, charging
and correlation in quantum dots~\cite{Aleiner02}, superconductivity or
ferromagnetism in metallic grains~\cite{VonDelft01}, and transport
through effectively one-dimensional systems~\cite{FisherGlazman97}.
Since interference effects are ubiquitous at the nanoscale, it is
natural to examine the interplay between interference and many-body
interactions in these systems.

A magnetic impurity coupled to an electron gas is a classic many-body
problem, known as the Kondo problem~\cite{Hewson93}.  In the simplest
case, one takes a spin-1/2 impurity coupled antiferromagnetically with
an exchange constant $J$ to an electron gas described by a constant
density of states $\rho_0$ within a bandwidth $D$.  The physics has
two distinct temperature regimes -- at high $T$ the
electrons scatter off the impurity in inelastic spin-flip processes,
while at low $T$ there is only elastic scattering as the impurity is
screened by electrons forming a bound state singlet. The crossover
takes place at an energy scale $T_K^0$, which in the weak coupling limit 
$\simeq De^{-1/J\rho_0}$. All
impurity properties (i.e. impurity contribution to thermodynamic
quantities and impurity correlators) have temperature dependencies
that are \textit{universal} after rescaling with $T_K^0$. This
simplest case corresponds to a bulk piece of metal in which there is
no interference. 

More recently, the effect of interference on Kondo impurities has been
studied in various open systems.  In disordered metals, the presence of
magnetic impurities affects the weak-localization properties
\cite{Okhawa83,Martin97}, and the fact that each impurity sees a
different local density of states may give rise to non-Fermi-liquid
signatures in the metal-insulator transition \cite{Dobrosavljevic92}.
In  quantum point contacts, Friedel oscillations associated with the
proximity of a boundary  lead to a coupling between the magnetic
impurity and the electron gas which depends on both the position of the
impurity and the energy of the electron state; as a consequence
fluctuations of the Kondo properties are observed \cite{Zarand96}. 

For closed systems, on the other hand, it is natural to ask: What is the
effect of the confinement on the many-body physics?  In ultra-small metal particles~\cite{VonDelft01}, for instance, the mean
single-particle level spacing $\Delta$ can be comparable to the bulk
Kondo temperature $T_K^0$. Clearly, a
magnetic impurity in such a particle will show new physical properties
associated with the new scale $\Delta$ produced by confinement. This
has been studied theoretically in both metallic
\cite{Thimm99} and semiconductor heterostructure
\cite{Simon02,Cornaglia03} systems.  However, most of these studies
have neglected the {\em fluctuations} that are inherent in any
mesoscopic sample (see however \cite{Kaul03,Lewenkopf05}).  Indeed, if
$\Delta$ is the microscopic scale induced by the confinement, this
latter brings in yet another energy scale, the Thouless energy $E_{\rm
Th}\simeq \hbar / t_{\rm f}$ where $t_{\rm f}$ is the time to cross
the nanostructure.  The range of energy between $\Delta$ and $E_{\rm
Th}$ is what is referred to as the mesoscopic regime.  For a magnetic
impurity coupled to electrons confined in a fully coherent
nanoparticle at temperatures within this mesoscopic range, both the
energies and the wave functions will display mesoscopic
fluctuations which are stronger than those in open systems.

The goal of this study is to understand the effect of these mesoscopic
fluctuations on the bulk universal behavior. Experimentally, our work
bears on two situations: first, an ultra-small metallic particle
\cite{VonDelft01} containing a single magnetic impurity, and, second,
the tunable Kondo impurities that have been formed using quantum dots
\cite{Goldhaber98b,cobden}.  We return in particular to the latter at
the end of the paper. Our approach is to combine analytic renormalization analysis \cite{Zarand96} with exact numerical results. The combination allows a quantitative understanding for $T \!\ge\! T_K^0$.

To describe a magnetic impurity coupled to a band of confined
electrons, we start with the single-particle electron states
$\epsilon_\alpha$ and wave functions $\phi_\alpha({\bf r})$ in the
absence of the impurity. The impurity is modeled as a localized state
at ${\bf r} \!=\! 0$ with onsite repulsion $U$. Finally, there is
hybridization between the impurity state $c_d$ and electron states
$c_\alpha$; for a $\delta$-like interaction, the electron gas is
completely described by the local density of states (LDOS) $\rho_{\rm
loc}(\epsilon) \!=\!$ $ \sum_{\alpha} | \phi_\alpha(0)|^2 \delta (\epsilon
- \epsilon_\alpha)$.

The combination of electron states that is most localized on the impurity is
$\psi_{\sigma}(0) \!=\! \sum_{\alpha} \phi_\alpha(0) c_{\alpha\sigma}$. The
impurity couples to only this state of the electron gas. With
locality and $SU(2)$ symmetry, the simplest model is
a symmetric (i.e.\ $\epsilon_d \!=\!$ $ -U/2$) Anderson model,
\begin{equation}  \label{anderson}
  H_{\rm And}=H_{\rm band}+  H_{\rm imp} + 
      V \sum_{\alpha \sigma} [\phi^{*}_{\alpha}(0) 
           {c^{\dagger}_{\alpha\sigma}}c_{d\sigma} + \mathrm{h.c.}]
\end{equation}
with $H_{\rm band} \!=\! \sum_{\alpha, \sigma}\epsilon_\alpha
c_{\alpha\sigma}^{\dagger} c_{\alpha\sigma}$ and $H_{\rm imp} \!=\!
\sum_{\sigma} \epsilon_d c_{d\sigma}^{\dagger} c_{d\sigma}+
Un_{d\uparrow} n_{d\downarrow}$.  We always take $U$ large enough so
that (1) is related through a Schrieffer-Wolff transformation to the
Kondo model
\begin{equation} \label{kondo}
H_{\rm Kondo}=H_{\rm band} + 
\sum_{\alpha \beta} J_{\alpha \beta}
\overrightarrow{S}\cdot c^{\dagger}_{\alpha\sigma_1}
\overrightarrow{\sigma}_{\sigma_1\sigma_2} c_{\beta\sigma_2}
\end{equation}
where $J_{\alpha \beta} \!=\! J\phi^{*}_{\alpha}(0) \phi_{\beta}(0)$ and
$J\!=\!8V^2/U$. The band in the Kondo model must be cut at an energy scale
$D_{\rm cut}$. ($D_{\rm cut}\!=\!0.18 \, U$ for the symmetric Anderson
model ~\cite{haldane,nrg}.) Numerical calculations will be performed
with the Anderson Hamiltonian Eq.~(\ref{anderson}), but most of our
discussion will involve the Kondo form (\ref{kondo}).

Our goal here is to understand the fluctuations in measured quantities
as the realization of the $\phi_\alpha(0)$ and $\epsilon_\alpha$
changes.  We focus on the case of chaotic confinement, for which a
random matrix theory (RMT) description of the one-body physics is
valid.  For definiteness, we study the local susceptibility,
 \begin{equation} \label{chiloc}
  \chi= \int_{0}^\beta d\tau \langle {S_z}^{d}(\tau){S_z}^{d}(0)\rangle
 \end{equation}
where ${S_z}^{d}(\tau)\!=\!e^{\tau H}\frac{1}{2}(n_{d \uparrow}-n_{d
\downarrow})e^{-\tau H}$. Experimentally, while this quantity is
inaccessible for magnetic impurities, we indicate in our concluding
discussion how to measure $\chi$ directly using quantum dots.  In
addition, most of our analysis and conclusions can be extended to
other properties.  In the absence of mesoscopic fluctuations
($\epsilon_\alpha$ equally spaced and $\phi_\alpha\!=\! {\rm const.}$) and
in the regime $\Delta\ll T \ll U$, $\chi$ shows universal behavior
\cite{Hewson93}: $T\chi\!=\!f(T/T_K^0)$ so that the energy scale $T_K^0$
is the only parameter.

\section{Numerics} 
Before turning to an analysis of the
Hamiltonian Eq.~(\ref{kondo}) using the ``poor man's scaling''
technique \cite{Anderson70}, we present exact numerical
data of $\chi$ for specific realizations of the closely related
Hamiltonian Eq.~(\ref{anderson}). A key point is that the non-perturbative numerical data
enables us to \textit{quantitatively verify} that the perturbative ``scaling''
approach is sufficient to capture many aspects of the mesoscopic
fluctuations, as well as to comment on the effect of these
fluctuations in the low temperature regime where perturbation-theory
is not valid. To get a non-perturbative hold on the Hamiltonian
Eq.~(\ref{anderson}), we use a Quantum Monte-Carlo (QMC) method due to
Hirsch and Fye~\cite{Hirsch86}. Because only the $d$-site in the
Anderson model is interacting, the introduction of one auxiliary
bosonic variable renders the entire action quadratic in the fermionic
fields. Integrating out the fermions leaves a determinant which is a
functional of the bosonic variable. Finally, using the determinant as
the weight to do importance sampling of the bosonic configuration, one
can calculate imaginary time correlators of the Anderson model. We
will study the correlator $\chi$ defined in Eq.~(\ref{chiloc}).

\begin{figure}[t]
\begin{center}
\includegraphics[width=3.2in,clip]{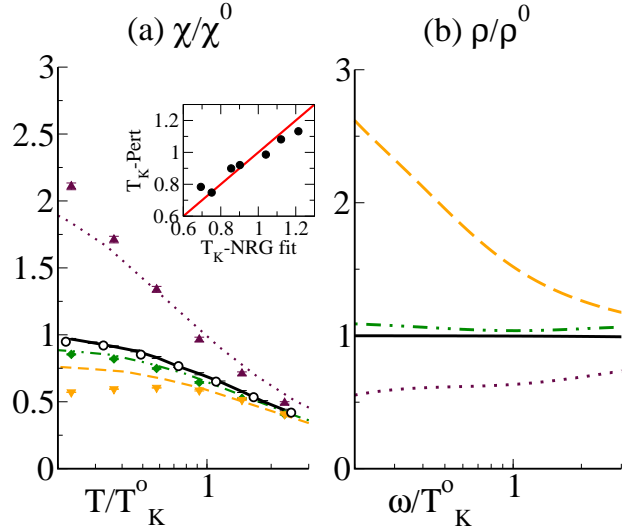}
\end{center}
\caption{(a) QMC data for the local susceptibility of
the Anderson model for the bulk-like case (open symbols) and three
realizations of a mesoscopic electron sea (filled symbols).  Also
shown are the corresponding universal curves $\chi\!=\!f(T/T_K)/T$ for a
realization dependent $T_K$ obtained from Eq.~(\protect \ref{eq:tkth})
(the correspondence between symbols and lines is the natural
one).  Inset: for realisations for which a good fit with a
universal curve is obtained, comparison of the $T_K^{\rm NRG}$
extracted from this fit , with the $T_K^{\rm pert}$ from perturbation
theory [Eq.~(\ref{eq:tkth})].  (b)~The smoothed local density of
states at the impurity site for the same three realizations and the
bulk case [line style matches the corresponding universal
curves in (a)]. Note the striking non-universal curves for
certain realizations and the correlation with variations in the
density of states.  The parameters are: $\Gamma \!\equiv\! \pi V^2/\Delta \!=\!125 \Delta ,U\!=\!250
\Delta$.  $T$ is scaled to the corresponding bulk Kondo
temperature $T_K^0 \!\simeq\! 10.2 \Delta$ (value from Bethe ansatz). The realization dependent
$\phi_\alpha$ and $\epsilon_\alpha$ are drawn from RMT-GOE and for the
bulk case are $\phi_\alpha\!=\!1$ with $\epsilon_\alpha$ equally spaced.  }
\label{fig:dos}
\end{figure}

Fig.~\ref{fig:dos}(a) shows $\chi(T)$ for four different realizations
of the conduction electron sea. The first one (open symbols)
corresponds to the ``bulk-like'' case, by which we mean a discrete
spectrum but without mesoscopic fluctuations (constant spacing of the
energy levels, no fluctuations of the wave functions).  The perfect
agreement with the universal curve demonstrates that, at the
temperature we consider, the discreetness of the spectrum per se plays
no role.\footnote{This implies the absence of any
even-odd effects such as the ones observed in \cite{Thimm99}, as well as
any difference between the grand canonical and canonical ensembles. At lower $T$ such effects would appear and could be captured by QMC techniques.}  
For the three
remaining curves (filled symbols), the one-body matrix describing the
conduction electron Hamiltonian is drawn from the Gaussian orthogonal
ensemble (GOE), which in particular implies independent Porter-Thomas
distributions for the intensities $|\phi_\alpha(0)|^2$.  Implicit here
is the assumption that $E_{\rm Th} \to \infty$, which, as discussed
below, is a well-defined limit in the chaotic case.
 
An important result arising from these calculations is that not all of
the QMC data can be described by the universal form,
$\chi\!=\!f(T/T_K)/T$, with a realization dependent $T_K$. This is
particularly evident in cases in which $\chi$ turns down at
low-$T$ since $f(T/T_K)/T$ is a monotonically decreasing function.  A
more detailed analysis shows that for the parameters used in
Fig.~\ref{fig:dos}, only about half of the realizations yield a temperature
dependent susceptibility that can be scaled onto the universal curve
with reasonably good accuracy.  {\em Clearly, mesoscopic fluctuations can cause
significant deviations from universality}.



\section{Poor Man's Scaling} 
The basic idea of this technique is
that Hamiltonian $H$ with bandwidth $D_{\rm cut}$ can be replaced with
another Hamiltonian $H^{\prime}$ with smaller bandwidth while keeping
the low energy physics unchanged. In the bulk case, the only coupling
parameter is $J\rho_0$, and its ``renormalization'' causes it to
diverge as the bandwidth approaches an energy scale that may be
identified with $T_K^0$. In our case of a discrete spectrum, we take
out a single level at a time. Following Anderson, upon removing the
topmost level $\phi_\beta$, in order to keep low energy physics
unchanged, the amplitude $J_{\alpha \gamma}$ between two low energy
states changes by $\delta J_{\alpha \gamma}\!=\!J_{\alpha
\beta}J_{\beta\gamma}/\epsilon_\beta $. Interestingly, this equation
is solved by $\delta J_{\alpha \gamma}$ of the form $(\delta
J)\phi^{*}_{\alpha}(0)\phi_{\gamma}(0)$ with
 \begin{equation}
    \delta J= J^2 \frac{|\phi_\beta(0)|^2}{\epsilon_\beta} \,,\quad
        \frac{\delta J}{\delta D}=J^2\frac{\rho_{\rm loc}(D)}{D} \,.
 \end{equation}
Thus in the ``poor man's scaling" approach, the effect of reducing $D$
is captured by letting $J$ flow, obtaining at low temperature $T$ the
effective coupling constant
 \begin{equation}  \label{eq:Jeff}
  J_{\rm eff}(T) \simeq
  \frac{J}{1 - J\int_{T}^{D_{\rm cut}} \rho_{\rm
  loc}(\epsilon) d \epsilon / \epsilon} \; ,
 \end{equation}
where $J$ is the ``bare'' coupling constant defined at the cutoff of
the Kondo model Eq.~(\ref{kondo}).  In the bulk problem it is well
known that the ``poor man's scaling'' result can also be reproduced by
two other perturbative approaches, namely a one-loop renormalization
group analysis, and Abrikosov's method for resumming the ``parquet''
diagrams~\cite{Abrikosov65}.  This is also true in the presence of
fluctuations (see e.g.~\cite{Zarand96} in the perturbative renormalization
group context). By working at finite temperature (using Matsubara
Green functions in conjunction with the other perturbative approaches)
$\rho_{\rm loc}(\epsilon)$ in Eq.~(\ref{eq:Jeff}) is replaced by a
more natural smoothed LDOS
\begin{equation} \label{eq:smooth}
  \rho_{\rm sm}(\omega) = \frac{\omega}{\pi} \int_{-\infty}^{\infty} d\epsilon 
  \frac{\rho_{\rm loc}(\epsilon)}{\omega^2 + \epsilon^{2}},
\end{equation}
In the following we shall discuss our results in terms of the smoothed
LDOS Eq.~(\ref{eq:smooth}).
We note that one-loop RG analysis has been shown to be a controlled approximation only for a flat (bulk) density of states; hence, for the mesoscopic regime, we turn to a comparison with the exact numerics.

\section{Comparison of Numerics with Poor Man's Scaling} 
At a qualitative level, the relevance of $\rho_{\rm sm}(\omega)$ can
be assessed by comparing panels (a) and (b) of Fig.~\ref{fig:dos},
where we have plotted the functions $\chi (T)$ and $\rho_{\rm
sm}(\omega)$ for the same realizations.  Note that the cases with
large deviations from the universal behavior are those for which
$\rho_{\rm sm}$ changes the most from the high to low energy
scales. The non-monotonic behavior of $\chi$ for one realization may
be understood as being due to a peak in $\rho_{\rm sm}$ at low
frequencies. In other cases a dip in $\rho_{\rm sm}$ causes $\chi$ to
increase continuously, perhaps until $T \!\sim\! \Delta$. In such cases
there is no ``Fermi-liquid'' regime ($\chi \sim$ constant and
$T>\Delta$) even though the bulk case has such a regime.

As the Kondo temperature is defined as the scale at which the
effective coupling constant $J_{\rm eff}$ diverges, it is natural to
evaluate the realization dependent {\em fluctuations} of $T_K$ around
the bulk value $T_K^0$ using the perturbative Kondo temperature
$T_K^{\rm pert}$ defined as
\begin{equation} \label{eq:tkth}
    \int_{T_K^{\rm  pert}}^{D_{\rm cut}} \rho_{\rm sm}(\omega) 
    \frac{d \omega }{ \omega}
    =
    \int_{T_K^0}^{D_{\rm cut}} {\overline{\rho}_{\rm sm}}(\omega) 
    \frac{ d \omega }{ \omega} \; ,
\end{equation}
where ${\overline{\rho}_{\rm sm}}$ is averaged over realizations and
so is by construction essentially constant.  The  lines in
Fig.~\ref{fig:dos}(a) correspond to the universal curves $\chi(T)
\!=\!f(T/T^{\rm pert}_K)/T$.  Clearly, the high temperature (i.e. above
$T_K^{\rm pert}$) part of the QMC data is well described in this way.
To evaluate how quantitatively Eq.~(\ref{eq:tkth}) describes the data,
we furthermore consider {\em all} our realisations that show
approximately ``universal'' behavior, in the sense that the data follow
the bulk universal curve obtained from NRG rescaled with an
appropriate fitted Kondo temperature. This constitutes approximately two-thirds of the realisations studied. We have then compared (inset in
Fig.~\ref{fig:dos}) this $T_K^{\rm NRG}$ {\em fit} with the energy
scale $T^{\rm pert}_K$ given by Eq.~(\ref{eq:tkth}). The agreement is
quite good, especially considering that Eq.~(\ref{eq:tkth}) is meant
to define an energy scale rather than an exact number.

It has to be borne in mind, however, that with the Hirsch and Fye
algorithm we are using, the ratio $U/T$ that can be practically
investigated cannot be made much larger than 500.  The parameters used
in Fig.~\ref{fig:dos} have been chosen in such a way that temperature
significantly smaller than the bulk Kondo temperature $T_K^0$ can be
reached, so that the absence of scaling behavior in the low
temperature regime could be demonstrated.  This however means that
$D_{\rm cut} \equiv 0.18 U \simeq 16.6 T_K^0$ is only
a little bit more than a decade larger than the bulk Kondo
temperature.  Here this implies that a good part of the
fluctuations seen in the high temperature part of Fig.~\ref{fig:dos}
could be explained by fluctuations of the smooth density of states
$\rho_{\rm sm}$ {\em at} $D_{\rm cut}$.

\begin{figure}[t]
\begin{center}
\includegraphics[width=3.0in,clip]{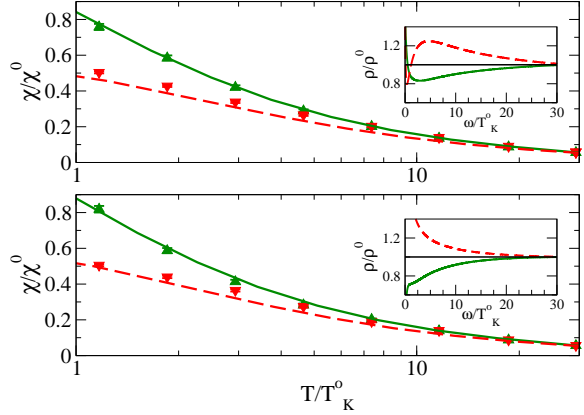}
\end{center}
\caption{Triangles: QMC calculation of the local
susceptibility for four realizations, with the parameters $\Gamma\!=\!250
\Delta$ and $U\!=\!1000 \Delta$, so that $D_{\rm cut} \!=\! 180 \Delta \simeq
35.3 T_K^0$, with  $T_K^0
\simeq 5.1 \Delta$.   
Lines: universal curves using $T^{\rm pert}_K$ from Eq.~(\ref{eq:tkth}).
Inset: Corresponding smoothed densities of states (with matching line styles).}
\label{fig:high_T}
\end{figure}

To show that the one-loop renormalization analysis indeed captures
the fluctuations associated with the energy dependence of the
density of states, we have performed another set of QMC
simulations, with the parameters $\Gamma\!=\!250 \Delta$ and $U\!=\!1000
\Delta$, so that $D_{\rm cut} \!=\! 180 \Delta$ and $T_K^0 \simeq 5.1 \Delta$.  The larger value of $D_{\rm
cut}$ already implies that $\rho_{\rm sm}(D_{\rm cut})$ shows
significantly less fluctuations.  To completely exclude the
possibility that the observed fluctuation of the susceptibility is an
effect of variations of the smooth density of states at the scale
$D_{\rm cut}$, we furthermore {\em rescale} all the density of states
[i.e.\ multiply the $\phi_\alpha(0)$ by a constant] so that $\rho_{\rm
sm}(D_{\rm cut}) \!=\! \rho_0$ for all realizations.  The resulting temperature
dependent susceptibility, for four realizations (selected because they
display large density of states fluctuations), is shown in
Fig.~\ref{fig:high_T}.  The comparison with the universal curves
$f(T/T^{\rm pert}_K)/T$ shows that in this case, also, the high
temperature part of the QMC data is well described in this way. It is particularly remarkable, in both Figs. 1 and 2, that the RG result works for temperatures as low as $T_K$ or even below.

We thus reach the central result of this paper: \textit{The excellent
agreement seen in Fig.~\ref{fig:high_T} between theoretical
predictions obtained in this way and exact numerical calculations
provides a numerical demonstration that this approach is a sound
starting point for including mesoscopic fluctuations.}

\section{Fluctuations of $T_{K}$} 
This agreement justifies using $T^{\rm pert}_K$ as the crossover
temperature from weak to strong coupling, even for realisations for
which the universal scaling is destroyed by the mesoscopic
fluctuations.  Within our RMT modeling, the variance of $T^{\rm
pert}_K$ can be easily evaluated from Eq.~(\ref{eq:tkth}) in the limit
$\delta T_{K} \!=\!$ $(T_{K}^{\rm pert} - T_K^0) \ll T_K^0$ as \cite{Kettemann04}
\begin{equation}
\label{eq:tkvar_rmt}
\frac{\langle {\delta T_{K}}^2  \rangle}{{{T}_K^0}^2} \simeq
\alpha\frac{\Delta}{{T}_K^0} \;.
\end{equation}
For our model, $\alpha\!=\! (4/\pi) \ln 2$; more generally it is a number
of order 1. An improved model can be developed through semiclassical
analysis, in which case $\langle \cdot \rangle$ may signify an energy
average.  The correlator $\langle \rho_{\rm sm}(\omega) \rho_{\rm
sm}(\omega') \rangle$ is then related to the Fourier transform of the
classical probability $P_{\rm cl}(t)$ that a particle starts at the
impurity with energy $E_F$ and returns there after time $t$
\cite{Argaman93}.  Eq.~(\ref{eq:tkvar_rmt}) can therefore be
generalized to
\begin{equation}
\label{eq:tkvar_sc}
\frac{\langle {\delta T_{K}}^2  \rangle}{{{T}_K^0}^2} \simeq
\frac{2 \Delta V}{\pi \hbar} \int_{T_K^0}^{D_{\rm cut}} \frac{d \omega'_1}{\omega'_1} 
\frac{d \omega'_2}{\omega'_2} \int_0^\infty dt P_{\rm cl}(t) 
e^{- ( \omega'_1 + \omega'_2) t/\hbar}  \; 
\end{equation}
($V$ is the volume of the dot).  
For chaotic systems, 
\begin{eqnarray} \label{eq:p_cl}
  P_{\rm cl}(t) = 0 \hspace{1.3em} & \quad {\rm for} & t <\hbar /  E_{\rm Th} \\
  P_{\rm cl}(t) = 1/V & \quad {\rm for} & t >  \hbar / E_{\rm Th} \;
\nonumber
\end{eqnarray}
so that Eq.~(\ref{eq:tkvar_rmt}) is then trivially recovered in the
limit $y \equiv E_{\rm Th}/T_K^0 \to \infty$.  The fluctuations of
$T_K$ vanish, of course, in the opposite limit $y \to 0$. A remarkable
implication of Eqs.~(\ref{eq:tkvar_sc})-(\ref{eq:p_cl}) is that the
value of $E_{\rm Th}$ does not affect the mesoscopic fluctuations for
a chaotic system with $E_{\rm Th} \gg T_K$.  The fluctuations are in
this sense ``universal'', and are described by RMT for any chaotic
system.

\section{Conclusions} 
We have shown that the mesoscopic fluctuations typical in
nanostructures significantly affect the physics of a Kondo impurity. Our 
main result is a demonstration that the {\em fluctuations} of the
high temperature properties can be understood by following a simple
poor man's scaling argument (or, equivalently, one-loop RG or
resummation of parquet diagrams) and that this works down to $T \!\approx\! T_K$.
In addition, we show 
(1)~deviations from universal behavior at low temperature,
(2)~an explicit formula for the variance of the Kondo temperature, and
(3)~a semiclassical connection between the fluctuations of $T_K$ and the classical probability of return.

Kondo physics has recently received renewed attention because of its
relevance to the low temperature physics of quantum dots
\cite{Goldhaber98b,cobden}.  These ``artificial magnetic impurities''
provide tunability of the microscopic Hamiltonian, and can be
naturally interfaced with larger quantum dots displaying the kind of
mesoscopic fluctuations that we address here.  From an experimental
point of view, the physics we are discussing should be easily
observable in such systems. In particular, we stress that though we
have here specifically considered the local susceptibility, the
qualitative features we have discussed -- the mere existence of the
mesoscopic fluctuations, the absence of universality of the low
temperature regime, and the universal character, but with a
realisation dependent Kondo temperature, of the $T \!\ge\! T_K$
regime -- should be observed for other physical quantities.  In the
same way, the fact that the the realisation dependent Kondo
temperature can be extracted from a fit to the universal form in the
high temperature regime implies that our quantitative predictions
Eqs.~(\ref{eq:tkvar_rmt})-(\ref{eq:tkvar_sc}) can also be tested on
physical quantities other than the local susceptibility.

Finally, we mention that experimental set-ups where the correlator
Eq.~(\ref{chiloc}) can be directly measured are also possible. In
particular, the problem of charge fluctuations in a gated, Coulomb
blockaded metal grain \cite{AshooriBermanPRL99} has been shown to be
equivalent to a Kondo problem~\cite{Matveev91}. In this mapping the
charge and gate voltage of the CB problem map to the spin of the Kondo
impurity and a \textit{local} magnetic field applied to it. The
differential capacitance at a degeneracy point in the CB problem is
thus equivalent to the \textit{local} susceptibility
Eq.~(\ref{chiloc}) of a Kondo problem. In the presence of a strong
polarizing magnetic field, the equivalent Kondo Problem is described
by exactly our Hamiltonian Eq.~(\ref{kondo}). Thus, a gated dot
connected to another dot provides a realization of a Kondo impurity
coupled to a mesoscopic electron gas where it is possible to measure
the correlator Eq.~(\ref{chiloc}) directly.

\acknowledgments

It is a pleasure to thank P.~Brouwer, A.~Finkelstein, L.~Glazman,
C.~Marcus, K.~Matveev, M.~Vojta, and J.~Yoo for valuable discussions.
This work was supported in part by the NSF (DMR-0103003).


\end{document}